\documentclass[conference]{IEEEtran1}
\usepackage[noadjust]{cite}

\ifCLASSINFOpdf
\usepackage[pdftex]{graphicx}
\else
\usepackage[dvips]{graphicx}

\fi
\usepackage[cmex10]{amsmath}
\usepackage{array}

\usepackage{xcolor}
\usepackage[tight,footnotesize]{subfigure}
\begin{document}
		%
		\title{High-Dimensional Quantum Key Distribution in Quantum Access Networks}
		
		\author{\IEEEauthorblockN{Osama Elmabrok,$^{1,^*}$ Mohsen Razavi,$^{2}$ Tawfig Eltaif,$^{3}$ and Khaled A. Alaghbari$^{4}$}
			\IEEEauthorblockA{\it \small$^{1}$Department of Electrical and Electronic Engineering, University of Benghazi, Libya\\$^{2}$School of Electronic and Electrical Engineering,
				University of Leeds, LS2 9JT, UK\\$^{3}$Faculty of Engineering Technology and Science, Higher Colleges of Technology, UAE\\$^{4}$Faculty of Engineering and Technology, Multimedia University (MMU), 75450, Malaysia\\ 
				\small$^{*}$osaelmabrok@gmail.com
			}
		}
	
	
	%


	\maketitle
	
	\begin{abstract}
		We investigate the use of high-dimensional quantum key distribution (HD-QKD) in wireless access to hybrid quantum-classical networks. We study the distribution of $d$-dimensional time-phase encoded states between an indoor wireless user and the central office on the other end of the access network. We evaluate the performance in the case of transmitting quantum and classical signals over the same channel by accounting for the impact of background noise induced by the Raman-scattered light on the QKD receiver. We also take into account the loss and background noise that occur in indoor environments as well as finite key effects in our analysis. We show that an HD-QKD system with $d=4$ can outperform its qubit-based counterpart. 
		
	\end{abstract}


	
	

	%
	\IEEEpeerreviewmaketitle

	
	\section{Introduction} 
	\label{sec:intro}
	Quantum key distribution (QKD) provides cryptographic security based on the laws of quantum mechanics rather than computational complexity~\cite{pirandola2020advances}. This can particularly be useful when long-term security is needed, as is the case in many scenarios where private data, such as medical records, may be shared via the Internet. As the common method for exchanging data by end users is via wireless access to communications networks, it would be useful for QKD systems offer their services in the wireless mode as well \cite{Quantum_Access_to_quantum_networks, GC2016, bahrani2019resource}. The presence of high levels of background noise in wireless channels, even in indoor environments~\cite{Wireless_indoor_QKD, Globecom15}, is, however, a significant challenge in wireless QKD systems~\cite{bouchard2018round}, resulting in low key rates. {The use of high-dimensional QKD (HD-QKD) protocols is an intriguing option as they are known to offer higher resilience to noise than their qubit-based counterparts~\cite{cozzolino2019high, PhysRevA.82.030301, ecker2019overcoming, bacco2019boosting}. This could particularly be useful, as we investigate here, for QKD systems that rely on wireless access to optical hybrid quantum-classical networks, i.e., where the optical access network is shared between quantum and classical applications.} 
	
	
	QKD has seen a rapid growth over the past few years. It has shown versatility by being demonstrated over various types of links, including optical fiber~\cite{frohlich2017long, QKD307km}, free space~\cite{steinlechner2017distribution, sit2017high}, satellite~\cite{ren2017ground, yin2017satellite}, and even underwater links~\cite{ji2017towards, bouchard2018quantum}. It has also been demonstrated over hundreds of kilometers~\cite{chen2020sending, boaron2018secure, yin2016measurement}, and in coexistence with classical channels over an optical fiber~\cite{kumar2015coexistence, wang2017long, eriksson2018coexistence}. The key generation rate, in all these cases, has to however increase for QKD to become competitive and commercially viable. 
	
	
	HD-QKD provides a promising and efficient platform for overcoming some of the practical issues faced by present QKD systems~\cite{PhysRevA.61.062308, cerf2002security, cozzolino2019orbital, wang2018multidimensional, kues2017chip, bavaresco2018measurements}. The efficiency is mainly due to its capacity of encoding multiple bits of information, $\textrm{log}_{2}(d)$ for $d$-dimensional system, on a single photon \cite{bao2016detector}, as well as its being more resilient to noise~\cite{cozzolino2019high}. The same degrees of freedom used in qubit-based QKD systems can be employed in a high-dimensional quantum state space. In fact, various degrees of freedom of the photon, such as time-energy and time-bin encoding~\cite{islam2017provably,  ali2007large}, path encoding~\cite{da2021path}, orbital angular momentum of light~\cite{cozzolino2019orbital, giordani2019experimental}, and frequency~\cite{kues2017chip, jin2016simple} have been demonstrated in HD-QKD systems. Each degree of freedom has its own benefits in terms of stability, control, and scalability, as well as its own challenges~\cite{cozzolino2019high}. 
	
	In this work, we examine the application of HD-QKD in quantum access networks. In our case, such networks are comprised of an indoor wireless link coupled to a fiber-based passive optical network (PON) catering both classical and quantum users. We employ a $d$-dimensional time-phase encoding, which is more compatible with current fiber networks, and analyse its performance in this setting. We analyse the effect of several sources of noise in our secret key analysis, such as the background noise caused by Raman-scattered light from classical channels. We also take into account the loss and background noise that occur in indoor environments~\cite{Wireless_indoor_QKD, Quantum_Access_to_quantum_networks, bahrani2019wavelength}. We obtain the key generation rate by accounting for finite-size key effects for the decoy state protocol~\cite{islam2017provably, lim2014concise}. Our results suggest that a simple HD-QKD setup with $d=4$ can already improve the reach and noise resilience of such QKD systems.
	
	The remainder of this paper is organized as follows. In Section 2, the system is described, with its key-rate analysis presented in Section 3. The numerical results are then discussed in Section 4, and Section 5 concludes the paper.

	\section{System Description} 
	
	\begin{figure}[t]
		\centering
		\includegraphics[width=.85\linewidth]{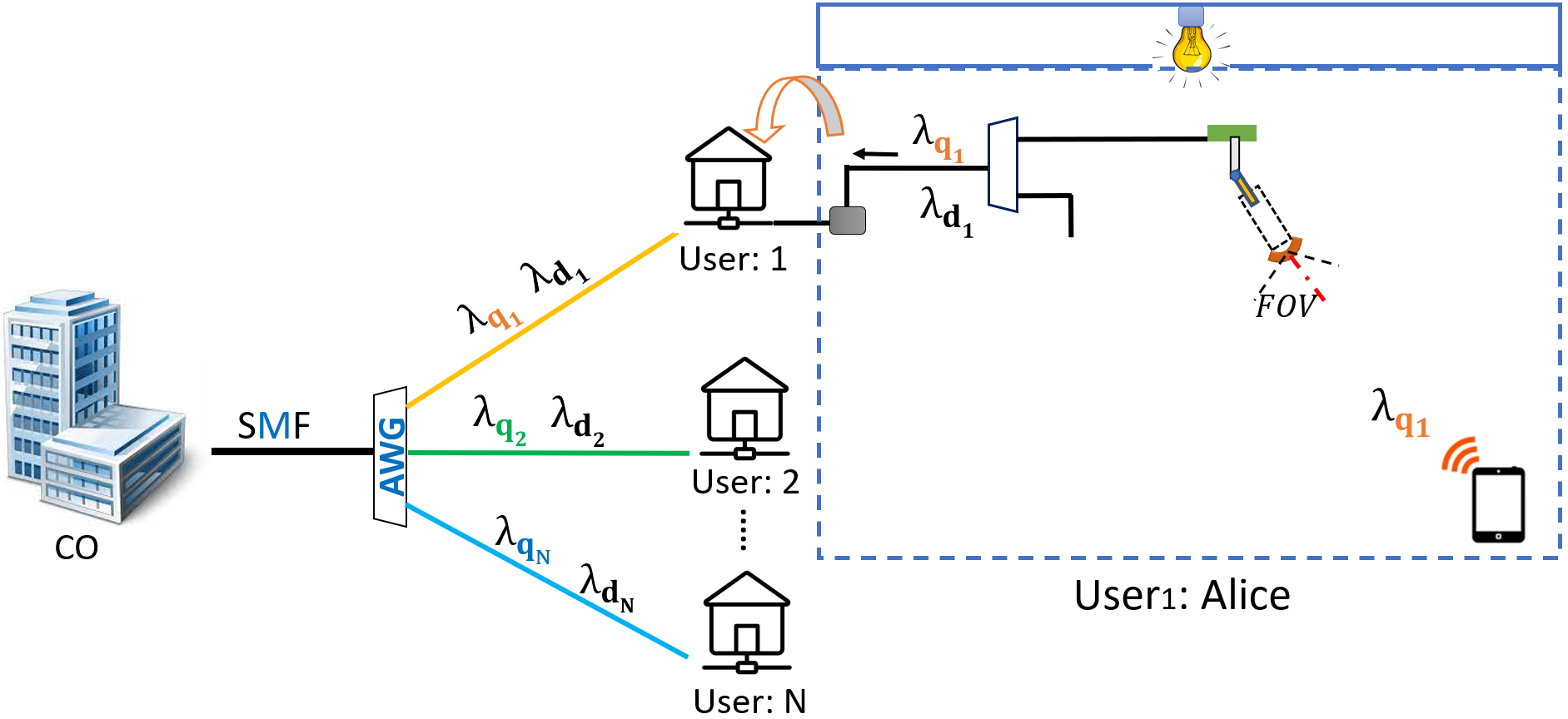}
		\caption{Schematic view of exchanging secret keys between an indoor wireless user, Alice, with a remote user, Bob. The latter is present at a central office at the end of an access network. SMF: Single-Mode Fiber, CO: Central Office, AWG: 
			Arrayed Waveguide Grating.}
		\label{fig_schematic_view}
	\end{figure}

	In this section, we describe the setup used to examine HD- QKD in hybrid quantum-classical access networks comprised of indoor optical wireless channels and single-mode optical fibers; see Fig.~\ref{fig_schematic_view}.  By utilising time-phase encoding technique, a block size of data, $N$, is sent from Alice to Bob.  Alice is assumed to be in a windowless room of size  $X \times Y \times Z$, lit by an artificial light source. On the room's floor, Alice uses a mobile device with a QKD transmitter that transmits light toward the ceiling, whereas Bob is assumed to be at the end of the access network in the central office. The signals sent by Alice are collected by a telescope, which is fixed at the center of the room's ceiling, and coupled to a single-mode fiber to be sent to the central office as shown in Fig.~\ref{fig_setup}. We assume that {$N_s$}  users are connected to the central office via a dense wavelength-division multiplexing (DWDM) PON to communicate both quantum, with wavelength $\lambda_{q_i}$ for user $i$, and classical, with wavelength $\lambda_{d_i}$, signals.  In our analysis, we focus on user 1 as the main example.
	
	In Fig.~\ref{fig_setup}, the telescope collects Alice's signals, which are then coupled into a single-mode fiber and forwarded to the central office where the measurements are performed. The wireless signals would suffer an extra coupling loss as a result of this coupling requirement. In this case, we assume that the telescope at the collection has a full alignment with the QKD source to ensure that a maximum power is received from the QKD source. This can be achieved by using additional beacon beams and MEMs steering mirrors \cite{chun2017handheld}.  In order to efficiently couple photons to the fibre, the effective field of view (FOV) at the collecting point should equal the numerical aperture of a single-mode fibre which would be roughly below $6^\circ$. In this setup, we use a $d$-dimensional time-phase encoding, which will be described in Sec.~\ref{subsection:Encoding}. In the following, we model the wireless and wired channels.

	\subsection{Channel Characterization}
	
	In this section, the two links in our setup, the wireless and fiber-based channels, are modelled in order to estimate the loss or background noise that may be introduced.
	
	\subsubsection{Indoor optical wireless channel}
	Loss and background noise can make it difficult for a QKD system to function. Path loss is often severe in indoor environments partly because of beam sperading and alignment requirements, but also, in our case, because of coupling issues to the optical fiber. that would shorten the key length. Background noise can also degrade the scheme's performance by increasing the error rate. Using optical wireless communications (OWC)  models, we estimate the path loss and background noise sneaked into the QKD receiver. The channel DC gain, $H_{\rm DC}$ \cite{kahn1997wireless,gfeller1979wireless}, which estimates how much of the transmitted power is collected by the telescope in the room, is used to model the channel transmittance for the wireless link. For the line-of-sight link between the QKD transmitter and receiver, this DC-gain is given by~\cite{kahn1997wireless}: 
	\begin{equation}
		\label{eq:HDC}   
		H_{\rm DC}= 
		\begin{cases} 
			\frac{A(m+1)}{2\pi d_i^2} \cos(\phi)^{m_{\textrm{dc}}} T_s(\psi)  \\
			\times  g(\psi) \cos(\psi),  ~~~~~~~~~~~~~~~~  0 \leq \psi \leq \Psi_c,  \\ 
			0 ~~~~~~~~~~~~~~~~~~~~~~~~~~~~~~~~~~ \text{elsewhere}, \end{cases}
	\end{equation}
	where $A$ is the effective area of the telescope and $d_i$ is the distance between the QKD source and the telescope; $\psi$ and $\phi$ are, respectively, the incidence angle with reference to the receiver axis and the irradiance angle characterizing the relative location and orientation of the transmitter and telescope modules; $T_s(\psi)$ is the optical filter transmission factor; $m_{\textrm{dc}}$ and $g(\psi)$ are, respectively, the Lambert's mode number used to identify the directivity of the source beam and the concentrator gain, given by
	\begin{equation}
		m_{\textrm{dc}}=\frac{-\ln 2}{\ln(\cos (\Theta_{1/2}))}
	\end{equation}
	and 
	\begin{equation}
		g(\psi)= \begin{cases} 
			\frac{n_i^2}{\sin^2(\Psi_{c})}, ~~~~~~ 0 \leq \psi \leq \Psi_c 
			\\ 
			0 ~~~~~~~~~~~~~~~~\psi > \Psi_c \end{cases},
	\end{equation}
	where $n_i$, $\Psi_c$, and $\Theta_{1/2}$ are, respectively, the refractive index of the concentrator, the telescope's FOV, and the semi-angle at half power of the light source. We neglect the reflected pulses from the walls, as they would arrive at a later time.
	
	In regard to estimating the background noise in our window-less room, we calculate such noise induced by the artificial lamp using OWC models~\cite{Wireless_indoor_QKD, Globecom15}. That would depend on the power spectral density (PSD) of the employed light source and the telescope's FOV. Depending on the latter, we can limit the amount of background noise that may sneak into the QKD receiver. Here, we account for the reflected light from the walls and the floor that would be collected at the telescope~\cite{Wireless_indoor_QKD, Globecom15}. 
	
	\subsubsection{Optical fiber link}
	
	As for the PON, we assume that arrayed waveguide grating (AWGs) are used to multiplex/demultiplex different wavelengths. The loss for each AWG is $\Lambda$ and loss coefficients of the fiber link are $\alpha$ and $\alpha_r$; see~Table~\ref{Table}. QKD channels in a fibre link that is carrying classical data are mostly affected by the background noise caused by Raman scattering. Due to strong classical signals, the Raman noise would spread over a wide range of frequencies, and that would result in populating the QKD receivers with unwanted signals~\cite{Eraerds2010_1Gbps}. Based on  their locations and the direction of light propagation, the receivers might be susceptible to forward and backward scattered light \cite{Bahrani2016orthogonal}, which are, respectively, given by~\cite{Eraerds2010_1Gbps,Patel2012coexistence}  
	\begin{align}
		\label{FO}
		{I^{f}_{R}}(I,L,\lambda_d,\lambda_q)=Ie^{-\alpha_r L}L \Gamma (\lambda_d,\lambda_q) \Delta \lambda 
	\end{align}
	and 
	\begin{align}
		\label{BA}
		{I^{b}_{R}}(I,L,\lambda_d,\lambda_q)=I\frac{(1-e^{-2 \alpha_r L })}{2 \alpha_r} \Gamma (\lambda_d,\lambda_q) \Delta \lambda.
	\end{align}
	In \eqref{FO} and \eqref{BA}, $I$ is the intensity of a classical signal at wavelength $\lambda_d$, and $\Delta \lambda$ is the optical bandwidth of the QKD receiver.  $L$ is the fiber length and $\Gamma (\lambda_d,\lambda_q)$ is the Raman cross section (per unit of fiber length and bandwidth), which can be measured experimentally. In this work, we have used the results reported in~\cite{Eraerds2010_1Gbps, Bahrani2016orthogonal} at $\lambda_d = 1550$~nm, so it can be adapted to any other wavelengths in the C band. We assume that the transmitted power $I$ is set to guarantee a bit error rate (BER) of no more than 10$^{-9}$ for all data channels.
	The QKD receiver would then detect a total average number of photons, induced by forward and backward scattering, respectively, given by 
	\begin{align}
		\label{murf}
		{\mu^{f}_{R}}=\frac{\eta_d {I^{f}_{R}} \lambda_q T_d}{hc}
	\end{align}
	and
	\begin{align}
		\label{murb}
		{\mu^{b}_{R}}=\frac{\eta_d {I^{b}_{R}} \lambda_q T_d}{hc}.
	\end{align}
	where $\eta_d$, $T_d$, and $h$, respectively, identify  detectors' quantum efficiency, their gate duration, and Planck's constant with $c$ being the speed of light in the vacuum.

	\begin{figure}[t]
		\centering
		\includegraphics[width=.85\linewidth]{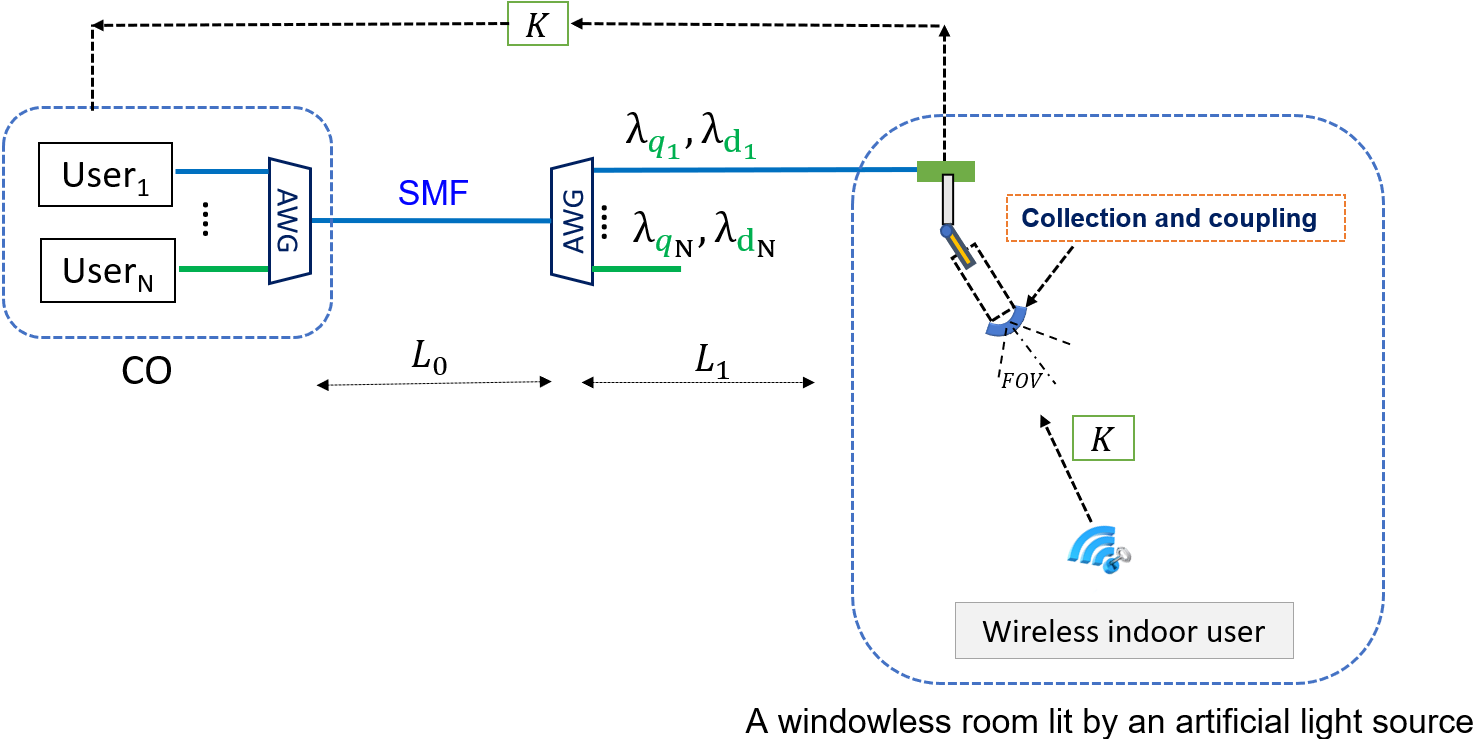}
		\caption{Secret keys are exchanged between Alice and Bob using HD-QKD. The QKD signals are collected and coupled to the fiber and sent to Bob, where the measurement is performed. Dynamic beam steering is used at the collection node. SMF: Single-Mode Fiber, CO: Central Office, AWG: Arrayed Waveguide Grating, K: the final key.}
		\label{fig_setup}
	\end{figure}

	\begin{figure}[t]
		\centering
		\includegraphics[width=.25\linewidth]{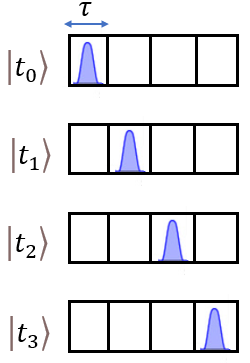}
		\caption{Temporal states for $d=4$ of time-phase encoding.}
		\label{fig_Temp_states}
	\end{figure}

	\subsection{High-dimensional time-phase encoding}
	\label{subsection:Encoding}
\begin{figure}[t]
		\centering
		\includegraphics[width=.85\linewidth]{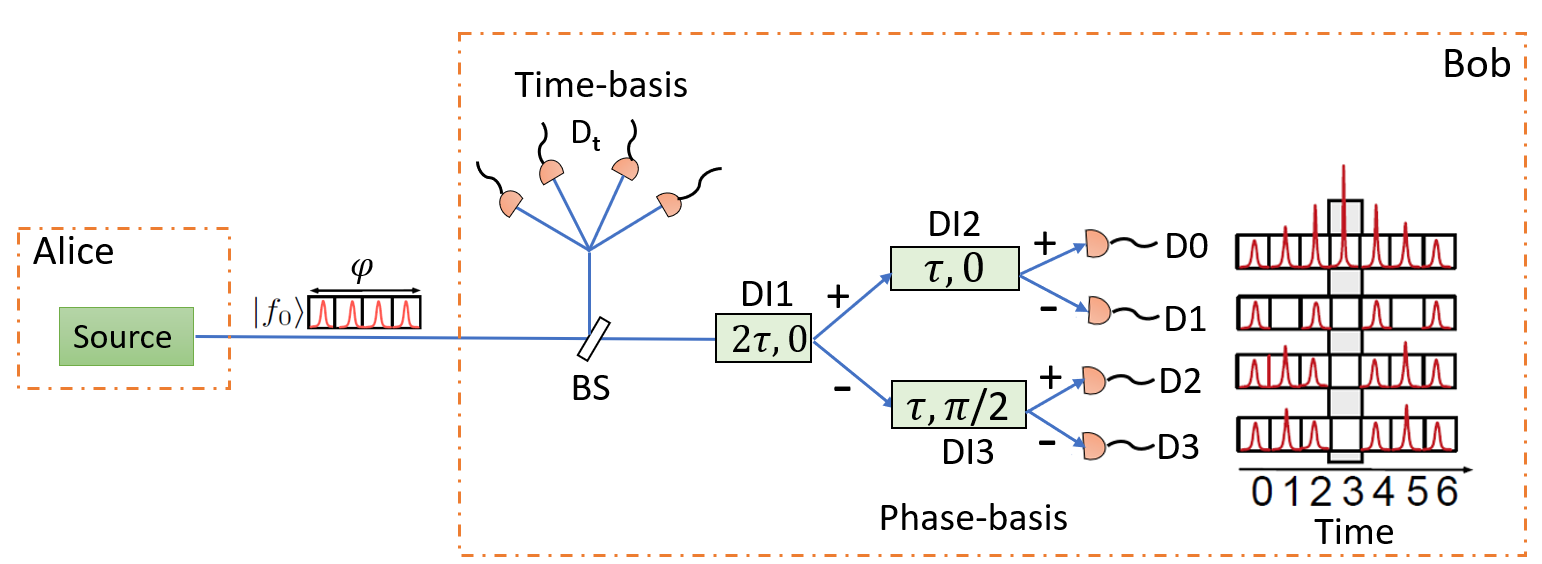}
		\caption{Measurement scheme for time-phase encoding technique for $d=4$. The received quantum states are randomly directed by a beamsplitter (BS) to either single-photon detectors or interferometers connected to single-photon detectors to measure the qudits in the time or phase bases, respectively. A tree of of time-delay interferometers are used to measure phase states. It consists of three time-delay interferometers (DIs) with their phases configured so that the input phase state ($|f_n\rangle$) and the detector in which the event is registered have a one-to-one mapping~\cite{islam2017provably}.   
		} 
		\label{fig_encoding}
	\end{figure}

	We assume that the time-phase encoding approach presented in~\cite{islam2017provably} is used for qudit encoding. In this technique, the sifted photon time-of-arrival data is used to compute the secret key, whereas phase measurement data is used  to monitor the presence of an eavesdropper. The quantum eigenstates in the time and phase bases are, respectively, denoted by  $\lvert t_m\rangle, m \in\{0, .., d-1\}$ and $\lvert f_n\rangle=\sqrt{d} \sum_{m=0}^{d-1}e^{2\pi inm/d} |t_m\rangle, n=0, ...,d-1$. $d$ temporal bins are assigned to each quantum state prepared in time or phase, for instance, Fig.~\ref{fig_Temp_states} shows temporal states for $d=4$. Each bin of width $\tau$ encodes $\textrm{log}_{2}(d)$ bits of information per photon. In Fig.~\ref{fig_encoding}, Alice sends out the quantum states over an untrusted quantum channel to Bob. The received quantum states are randomly directed by a beamsplitter to either single-photon detectors or interferometers connected to single-photon detectors to measure the qudits of $d=4$ in the time or phase bases, respectively. 
	
	
	In our key-rate analysis, we assume that the interferometric measurement method is used where a tree of time-delay interferometers is used to measure the phase states. We assume that the time basis, T, is chosen with probability $p_T$, whereas the phase basis, F, is chosen with probability $p_F=1-p_T$. Key-rate analysis is provided in the following section using a three-intensity decoy state method with finite-size effects.
	
	\section{key-Rate Analysis}
	
	The secret key length for HD-QKD using the decoy-state technique, with finite-size effects, is given by~\cite{islam2017provably, lim2014concise}:
	\begin{align}
		\label{KR}
		l=&\underset{\beta\in(0, \frac{\epsilon_{sec}}{22})}{\textrm{max}}\bigg[ \textrm{log}_{2}(d).\tilde s_{T, 0} + \tilde s_{T, 1}[c_i-h_d(\lambda ^U)]-  
		\textrm{leak}_{EC} & \nonumber \\
		&-\textrm{log}_{2}\frac{32}{\beta^8 \epsilon_{cor}}\bigg],
	\end{align}
	where $\tilde s_{T, 0}$ and $\tilde s_{T, 1}$ are the number of vacuum and single-photon detection events, respectively, in the time basis; $\epsilon_{cor}$ and $\epsilon_{sec}$ are the security parameters of the protocol corresponding, respectively, to error correction and privacy amplification parts; $c_i$ is the overlap parameter between the two used bases, time and phase, defined by $c_i:= -\textrm{log}_{2}\underset{n, m}{\textrm{max}}\mid\langle f_n\lvert t_m\rangle\mid^2$. In our key rate analysis we assume that $c_i=\textrm{log}_{2}(d)$; $\lambda ^U$ is the upper bound on phase error rate in single photons in the phase basis, and $h_d(\lambda ^U)$ is the Shannon entropy in the $d$-dimensional case, given by:
	$h_d(x) :=-x\textrm{log}_{2}(x/d-1)-(1-x)\textrm{log}_{2}(1-x)$. The number of bits sacrificed during error correction in \eqref{KR} is computed by {$\textrm{leak}_{EC} = f h_d(e_T) n_T$}, where $f$ is the error correction inefficiency, $n_{T}$ and $e_T$ are, respectively, the number of signals and the average error rate in the time basis.  The latter is calculated by $e_{T}=\sum\limits_{k=1}^3 p_{\mu_{k}} e_{T_{k}}$, where $p_{\mu_{k}}$ is the probability of using intensity (mean photon number) $\mu_k$, and $e_{T_{k}}$ is its corresponding error rate in the time basis. Appendix~\ref{App:KeyRate} shows how each of the above parameters can be calculated in the simulation scenario where no eavesdropper is present.
	
	In practical settings, the secret key length $l$ mainly depends on the overall transmissivity of each link, $\eta$, and the total background noise denoted by $n_N$. In Fig.~\ref{fig_setup}, the total Raman noise power for forward and backward scattering, denoted by $I^{f}_{T}$ and $I^{b}_{T}$ are, respectively, given by
	\begin{align}
		I^{f}_{T} =& [I_R^f(I,L_0 + L_1,\lambda_{d_1},\lambda_{q_1}) \nonumber \\
		&+ \sum_{u=2}^{N_s} {I_R^f(Ie^{-\alpha_r L_k},L_0,\lambda_{d_k},\lambda_{q_1})}] 10^{-2\Lambda/10}
	\end{align}
	and
	\begin{align}
		I^{b}_{T} = & [I_R^b(I,L_0 + L_1,\lambda_{d_1},\lambda_{q_1}) 
		\nonumber \\
		&+ \sum_{u=2}^{N_s} {I_R^b(I,L_0,\lambda_{d_k},\lambda_{q_1})}] 10^{-2\Lambda/10},
	\end{align} 
	where $L_0$ is the total distance between the AWG box at the users' side and the central office. $L_u$ is the distance of the $u$th user to the same AWG in the access network.  The total background noise at the Bob's end is then given by 
	\begin{align}
		\label{tot_backgroundNoise}
		n_N = \frac{ \eta_{d}\lambda_{q_1} T_d}{hc} \left( {I^{f}_{T}} + {I^{b}_{T}}\right)  + \eta_{d}n_{B}\eta_{\rm fib} \eta_{\rm coup}
		+ n_{dc},
	\end{align}
	where $\eta_{\rm coup}$ is the additional air-to-fiber coupling loss. $n_B$~\cite{Quantum_Access_to_quantum_networks, Wireless_indoor_QKD} and $n_{dc}$ are, respectively, the indoor background noise induced by the bulb and dark count rate; $\eta_{\rm fib}$ is the optical fiber channel transmittance including the loss due to AWGs. The total channel transmittance between the QKD transmitter and receiver is given by $\eta=H_{\rm DC}\eta_{\rm coup}\eta_{\rm fib}\eta_{d}$, where $\eta_{\rm fib}$ is given by $\eta_{\rm fib}=10^{-[\alpha(L_1+L_0)+2\Lambda]/10}$.
	
	\section{Numerical Analysis}
	\label{Sec:NumericalResults}
	
	\begin{table}
		\caption{Nominal values used for our system parameters.}      
		\begin{tabular}{|c|c|} \hline
			Parameter & Nominal value\\ \hline
			Room size, $X$,$Y$,$Z$ & ($4 \times 4 \times 3$) m$^3$ \\
			Semi-angle at half power of the bulb, $\Theta_{1/2}$, & $70^{\circ}$ \\
			Semi-angle at half power of the light source, $\Phi_{1/2}$  &   $1^{\circ}$ \\
			Receiver field of view, FOV  &   $6^{\circ}$ \\
			Time gate duration, $T_{d}$   & 100 ps \\ 
			$\mu_1$, $\mu_2$, $p_{\mu_1}$, $p_{\mu_2}$, $p_{T}$  & $0.54$, $0.1$, $0.5$, $0.06$, $0.9$\\
			$\mu_3$, $p_{\mu_3}$   & 0.0002, 0.44\\
			the air-to-fiber coupling loss, $\eta_{\rm coup}$ & 10 dB \\
			Fiber attenuation coefficient, $\alpha$, $\alpha_r$   & 0.2 dB/km, 0.046 /km \\
			Loss due to each AWG, $\Lambda$ & 2 dB \\
			Error correction inefficiency, $f$   & 1.16 \\
			Number of users, $N_s$ & 32 \\
			Dark count rate, $n_{dc}$   & $10^{-7}$ {ns}$^{-1}$ \\ 
			Misalignment probability, $e_{d}$   & 0.033 \\    
			Quantum efficiency of detector, $\eta_{d}$ & 0.3 \\
			\hline  
		\end{tabular}
		\label{Table} 
	\end{table}

	In this section, we provide some numerical results for the setup shown in Fig.~\ref{fig_setup}. We consider a DWDM setting with 0.8~nm channel spacing in the C-band with 32 users.  We assume  that $\lambda_{q_1}$ is 1555.62 nm and the corresponding $\lambda_{d_1}$ is 1585.2 nm. We consider a variable launch power for the classical channels to reduce the impact of Raman noise~\cite{ Quantum_Access_to_quantum_networks}. This corresponds to a -38.5 dB receiver sensitivity, ensuring a BER of $10^{-9}$. We assume that the fibre length from Alice's location to AWG, $L_1$, is 500 metres, and that this value is the same for all users. We assume that a full alignment is achieved between the QKD source and the telescope on the room ceiling in order to improve channel transmittance in the room. In this case, we assume that the semi-angle at half power of the QKD source, which is placed at a corner of the room’s floor, is 1$^{\circ}$ and the telescope's FOV is 6$^{\circ}$. We compute the key generation rate with finite-size effects~\cite{islam2017provably} using the decoy state approach of three intensities, $\mu_1$, $\mu_2$, and $\mu_3$. Table~\ref{Table} summarises the other nominal parameter values used in our simulation. 
	
\begin{figure}[t]
		\centering
		\includegraphics[width=.85\linewidth]{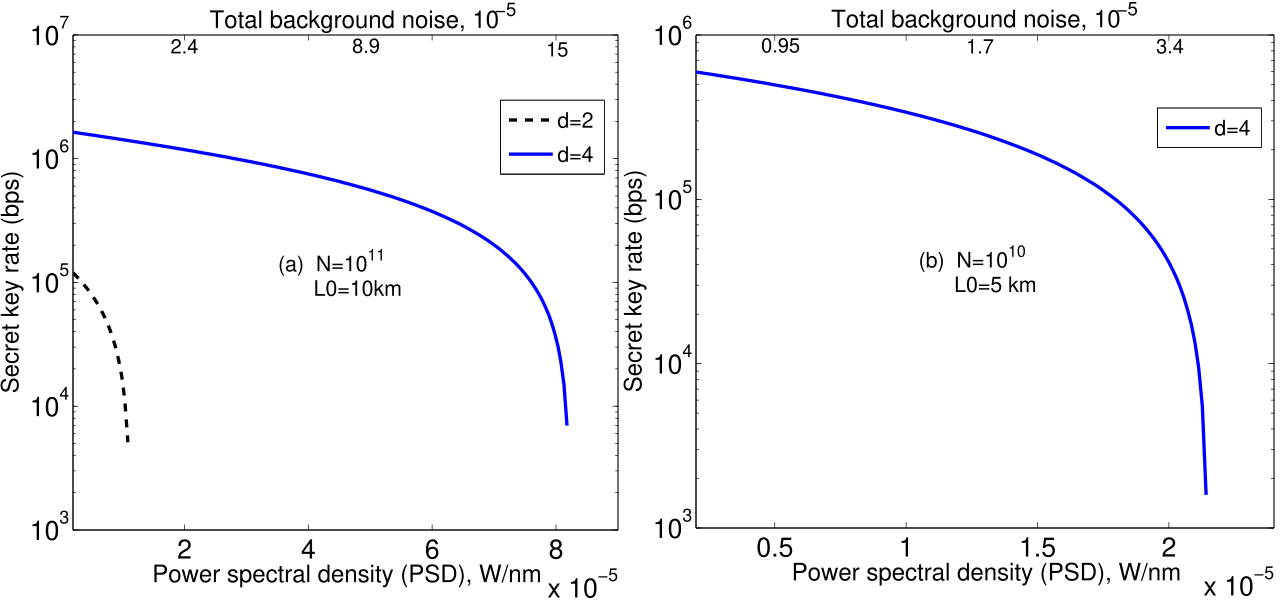}
		\caption{The secret key rate (bps) versus the power spectral density (PSD) of the bulb when the clock rate is 2.5 GHz. The top x-axis shows the corresponding total background noise, $n_N$. Eqs \eqref{nTK}-\eqref{mFK} in Appendix~\ref{App:KeyRate} show the expected values of background noise used to compute the secret key rate.  We consider a block size of data $10^{11}$ in (a) and $10^{10}$ in (b). The total channel transmittance is computed by assuming that a full alignment is attained between the QKD transmitter and the telescope where $\Phi_{1/2}$ = $1^{\circ}$ and FOV = $6^{\circ}$. We assume that the transmittance of the Mach-Zehnder interferometer, $\eta_{i}$, in (a) is 0.5 (3 dB loss) whereas it is approaching 1 ($\approx$ 0 dB loss) in (b). $L_0$ is assumed to be 10 km and 5 km in (a) and (b), respectively. 
		}
		\label{fig_rate_vs_PSD}
	\end{figure}

	One main challenge of using OWC for QKD purposes is the existence of severe ambient light, which affects the scheme’s performance, particularly due to artificial light sources. In Fig.~\ref{fig_rate_vs_PSD},
	the performance is assessed by varying the bulb's PSD and computing the corresponding secret key rate (bps) when the clock rate is 2.5 GHz~\cite{islam2017provably}. The estimated total background noise due to the bulb's PSD is shown on the top x-axis in Fig.~\ref{fig_rate_vs_PSD}.  The more levels of background noise in the channel means more errors would be generated, and, accordingly, at some point, no secret keys can be exchanged. The secret key rate is computed for $d=2$ and $d=4$ considering $N= 10^{11}$ and $N= 10^{10}$.  Figure~\ref{fig_rate_vs_PSD}(a) shows the improvement in boosting the key rate and tolerating more background noise when employing HD-QKD, $d=4$, in comparison to $d=2$ for $N= 10^{11}$ at $L_0=10$ km. With $d=4$, one can achieve a reasonable key rate using a simple QKD receiver with a few number of interferometers; see Fig.~\ref{fig_encoding}. In Fig.~\ref{fig_rate_vs_PSD}(b), if Alice and Bob exchanged a block size of data of $N= 10^{10}$ when $d=4$, they could exchange secret keys as long as the fiber length is less than 9~km, and we assume it to be 5~km. However, for $d=2$ the key rate cannot be extracted due to increasing the statistical errors in addition to the background noise. Interestingly, using high-dimensional quantum states, $d=4$, would allow for more background noise to be tolerated while still obtaining reasonable key rates. 
	
\begin{figure}[t]
		\centering
		\includegraphics[width=.85\linewidth]{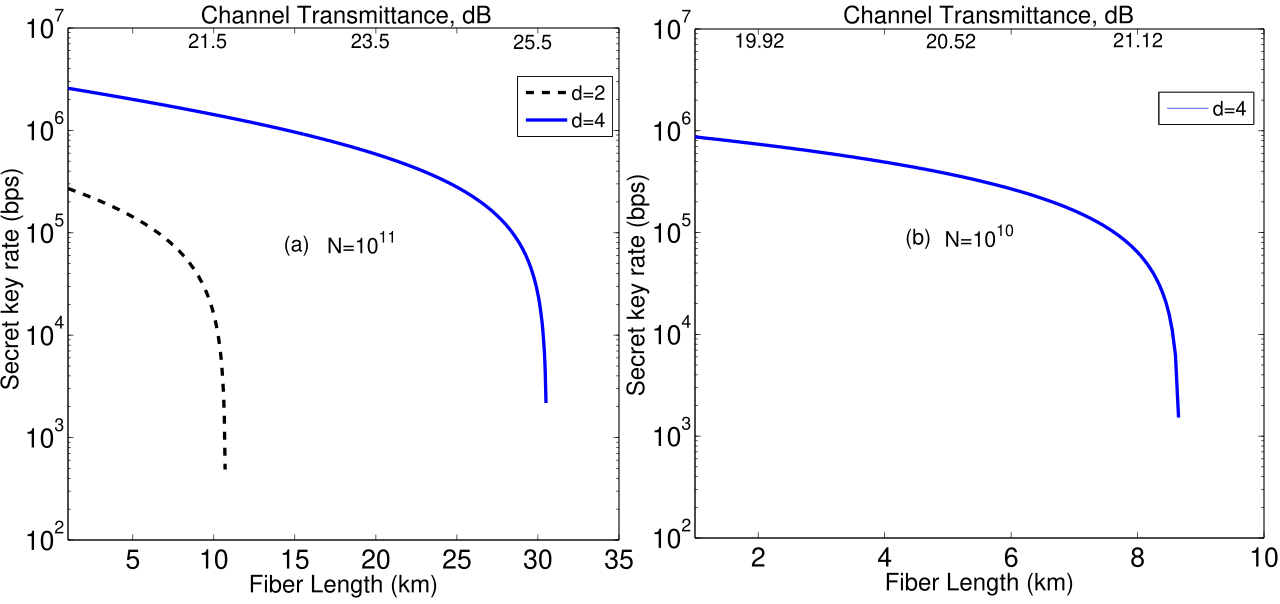}
		\caption{The secret key rate (bps) versus the fiber length ($L_0 + L_1$). The top x-axis shows the total channel transmittance, $\eta$. We consider a block size of $10^{11}$ in (a) and $10^{10}$ in (b). The total channel transmittance is computed by assuming that a full alignment is attained between the QKD transmitter and the telescope where $\Phi_{1/2}$ = $1^{\circ}$ and FOV = $6^{\circ}$. We assume that the transmittance in Mach-Zehnder interferometer, $\eta_{i}$, in (a) is 0.5 (3 dB loss), whereas it is approaching 1 ($\approx$ 0 dB loss) in (b). In both figures, we assume that PSD is $10^{-5}$ W/nm. 
		}
		\label{fig_rate_vs_length}
	\end{figure}

	Figure~\ref{fig_rate_vs_length} shows the secret key rate in bps versus the fiber length ($L_0 + L_1$), where the total channel transmittance in dB is shown on the top x-axis. This includes the coupling loss, $\eta_{\rm coup}$, which is 10 dB here; see Table~\ref{Table} for all loss parameters. In Fig.~\ref{fig_rate_vs_length}(a), the key rate is improved by roughly one order of magnitude when $d=4$ for $N= 10^{11}$, in addition to the transmission distance enhancement. Using HD-QKD of $d=4$ in this case would allow users to access PON networks without distance limitations in comparison to $d=2$. The case in Fig.~\ref{fig_rate_vs_length}(b) when $N= 10^{10}$, is more practical for applications that need acquisition time of about 4 sec when the clock rate is 2.5 GHz. For instance, when $d=4$ at 5 km where the total loss is 20.52 dB, one can obtain a secure key of length $3.7\times10^{5}$ bps, whereas no secret key can be obtained at $d=2$. 
	
	The simplest methods for generating high-dimensional quantum states are time-energy and time-bin encoding. However, as the dimensions of quantum systems increase, there are other practical considerations that we need to account for. For instance, the repetition rate of the generated states may decrease quickly, or the complexity of the setup may increase. That could be a significant concern for certain applications~\cite{da2021path}. The required dimension is determined by the use-case scenario. For instance, in Fig.~\ref{fig_rate_vs_length}(a), at 10 km, with the clock rate of 2.5 GHz, $d=4$ is sufficient for obtaining a secure key rate of about 1.4 Mb/s, which can be used to realize one-time pad encryption of video data~\cite{TokyoQKDNetwork2011}.

	\section{Conclusion} 
	
	High-dimensional QKD systems are interesting options for boosting error tolerance and for improving the secret key generation rate. In this work, we examined HD-QKD in quantum access networks considering the challenges of background noise, Raman noise, and path loss. We considered a finite size of data to exchange secret keys between a wireless indoor user and a remote user located at the central office. HD-QKD turned out to be a promising solution for users in indoor environments who might want to use applications such as video conferencing in a convenient and safe manner. For such applications, an HD-QKD system with dimension four could be sufficient for obtaining a reasonable secure key rate over PONs.  
	
	\vspace{+.2in}

	\vspace{-.2in}

	\section*{Acknowledgment}
	
  All data generated in this paper can be reproduced by the provided methodology and equations.
	
	\appendices
	\section{Key rate analysis}
	\label{App:KeyRate}

	In this appendix, the relevant parameters for calculating the secret key generation rate, under the nominal condition of no eavesdropping, are obtained. In \eqref{KR}, $\tilde s_{T, 0}$ and $\tilde s_{T, 1}$ are the number of vacuum and single-photon detection events, respectively, in the time basis, and they are given by~\cite{islam2017provably} and~\cite{lim2014concise}:
	\begin{align}
		\label{S_T0}
		\tilde s_{T, 0}= {\textrm{max}}\bigg[   \bigg( \frac{\tau_{0}}{\mu_2-\mu_3}\Big(\frac{\mu_2e^{\mu_3}n^{-}_{T_,\mu_3}}{p_{\mu_3}}-\frac{\mu_3e^{\mu_2}n^{+}_{T_,\mu_2}}{p_{\mu_2}} \bigg), 0   \bigg], 
	\end{align}
	and
	\begin{align}
		\label{S_T1}
		\tilde s_{T, 1}=& {\textrm{max}}\Bigg\{ \frac{\mu_1\tau_1}{\mu_1(\mu_2-\mu_3)-(\mu^2_2-\mu^2_3)} 
		\times \bigg[\frac{e^{\mu_2}n^{-}_{T_,\mu_2}}{p_{\mu_2}} 
		\nonumber \\  &-  \frac{e^{\mu_3}n^{+}_{T_,\mu_3}}{p_{\mu_3}}
		+ \frac{\mu^2_2-\mu^2_3}{\mu^2_1}\bigg(\frac{\tilde s_{T, 0}}{\tau_{0}}- 
		\frac{e^{\mu_1  n^{+}_{T_,\mu_1}}}{p_{\mu_1}} \Big)\bigg], 0 \Bigg\}, 
	\end{align}
	where $n^{\pm}_{T ,k}$ are the number of detections observed by Bob when Alice encodes the quantum states in the time basis with an intensity, $k$, and they are given by $n^{\pm}_{T ,k} = n_{T ,k} \pm \delta(n_T,\beta)$. The deviation term is assumed to be $\delta(n_T,\beta)=\sqrt{n_T/2  \textrm{log}(1/\beta)}$, following the Gaussian approximation (which is known to be closely following the more accurate bounds based on Chernoff inequality), for a total detection of $n_T$. $p_{\mu_k}$ is the probability that Alice sends a pulse with an average number of photons $\mu_k$ and $\tau_{n}$ is the probability that Alice sends an $n$-photon state, given by $\tau_{n}:= \sum\limits_{k\in K} e^{-k} \frac{K^n p_k}{n!}$ for $K=\Big\{ \mu_1, \mu_2, \mu_3 \Big\}$.
	
	In \eqref{KR}, $\lambda^U$ is an upper bound on the phase error rate of single photons in the phase basis, and it is given by $\lambda^U=\tilde Q +\xi$, where $ \tilde Q = \frac{\tilde \nu_{F, 1}}{\tilde s_{F, 1}}$ and $\xi = \sqrt{\frac{(\tilde s_{T, 1}+\tilde s_{F, 1})(\tilde s_{F, 1}+1)}{\tilde s_{T, 1}(\tilde s_{F, 1})^2}\log \frac{2}{\beta}}$. In $\tilde Q$ and $\xi$, $\tilde s_{F, 1}$ is the number of detected signal photon states in the phase basis, and it is estimated as in \eqref{S_T1} by replacing $T$ with $F$ for the corresponding phase-basis parameters. $\tilde \nu_{F, 1}$ is the number of erroneously detected single photons in the phase basis, and it is given by     
	\begin{align}
		\tilde \nu_{F, 1} = \frac{\tau_1}{\mu_2-\mu_3}(\frac{e^{\mu_2}m^{+}_{F_,\mu_2}}{p_{\mu_2}}-\frac{e^{\mu_3}m^{-}_{F_,\mu_3}}{p_{\mu_3}}), 
	\end{align}
	where $m^{\pm}_{F ,k} = m_{F ,k} \pm \delta(m_F,\beta)$, are the number of errors associated with the phase basis.
	
	In above equations, $n_{T ,k}$, $n_{F ,k}$, $m_{T ,k}$, and $m_{F ,k}$ are, respectively, given by
	\begin{align}
		\label{nTK}
		n_{T,k} = p_{\mu_k} p_T{^2}N(1-e^{-\eta \mu_k} +n_N), 
	\end{align}
	\begin{align}
		\label{nFK}
		n_{F,k} = p_{\mu_k} p_F{^2}N(1-e^{-\eta \eta_{i} \mu_k} +\frac{(d-1)}{d}n_N), 
	\end{align}
	\begin{align}
		\label{mTK}
		m_{T,k} = p_{\mu_k} p_T{^2}N(e_d(1-e^{-\eta \mu_k})+\frac{(d-1)}{d}n_N), 
	\end{align}
	and
	\begin{align}
		\label{mFK}
		m_{F,k} = p_{\mu_k} p_F{^2}N(e_d(1-e^{-\eta \eta_{i} \mu_k})+\frac{(d-1)}{d}n_N), 
	\end{align}
	where $p_F$ and $p_T$ are, respectively, the probability of encoding in the phase and time bases; $n_N$ is the total background noise; $e_d$ is misalignment error probability; $\eta$ is the channel transmittance.



	
	
	\bibliographystyle{IEEEtran}
	\bibliography{Master1}
	%
	
	
	

\end{document}